\documentclass[graybox]{svmult}
\usepackage[english]{babel}
\usepackage[utf8x]{inputenc}
\usepackage[T1]{fontenc}
\usepackage{amsmath}
\usepackage{graphicx}
\usepackage[usenames,dvipsnames]{xcolor}
\usepackage[colorinlistoftodos]{todonotes}
\usepackage[colorlinks=true,allcolors=blue]{hyperref}
\usepackage{listings}
\usepackage{todonotes}

\usepackage{mathptmx}      
\usepackage{helvet}        
\usepackage{courier}        
\usepackage{type1cm}                               
\usepackage{makeidx}        
\usepackage{graphicx}                                     
\usepackage{multicol}        
\usepackage[bottom]{footmisc}

\definecolor{color:keyword}{rgb}{0.53,0.05,0.05}
\definecolor{color:comment}{rgb}{0.25,0.37,0.75}
\definecolor{color:string}{rgb}{0.87,0.0,0.0}
\usepackage{bold-extra}

\lstdefinelanguage{Jolie}{
morekeywords={
	provide,until,OneWay,RequestResponse,new,
	main,define,inputPort,outputPort,init,execution,include,
	cset,if,else,csets,interface,type,throws,global,constants,for,
foreach,while,int,double,raw,void,undefined,string,long,bool,any,single,
sequential,concurrent,Jolie,Java,JavaScript,embedded,Location,Protocol,
Interfaces,Aggregates,scope,install,cH,comp,throw,this,default,synchronized,
nullProcess,false,true
},
sensitive=true,
morecomment=[l]{//},
morecomment=[s]{/*}{*/},
morestring=[b]",
otherkeywords={;,|,:}
}

\makeindex  

\begin{document}

\title*{Microservices: a Language-based Approach}
\author{Claudio Guidi, Ivan Lanese, Manuel Mazzara, Fabrizio Montesi}
\institute{Claudio Guidi \at italianaSoftware srl, Imola, Italy \email{cguidi@italianasoftware.com}
\and Ivan Lanese \at Focus Team, University of Bologna/INRIA, Italy \email{ivan.lanese@gmail.com}
\and  Manuel Mazzara \at Innopolis University, Russian Federation \email{m.mazzara@innopolis.ru}
\and Fabrizio Montesi \at University of Southern Denmark
\email{fmontesi@imada.sdu.dk}
}

\maketitle

 \abstract{Microservices is an emerging development paradigm where software is obtained by composing autonomous entities, called (micro)ser\-vices. However, microservice systems are currently developed using gene\-ral-purpose programming languages that do not provide dedicated abstractions for service composition.
Instead, current practice is focused on the deployment aspects of microservices, in particular by using containerization.
In this chapter, we make the case for a language-based approach to the engineering of microservice architectures, which we believe is complementary to current practice. We discuss the approach in general, and then we instantiate it in terms of the Jolie programming language.}

\section{Introduction}

Microservices~\cite{F14,N15,Dragoni2017} is an architectural style stemming from Service-Oriented Architectures (SOAs)~\cite{mackenzie2006}. Its main idea is that applications are composed by small independent building blocks -- the (micro)services -- communicating via message passing. Recently, microservices have seen a dramatic growth in popularity, both in terms of hype and of concrete applications in real-life software~\cite{N15}. Several companies are involved in a major refactoring of their backend systems~\cite{DDLM2017} in order to improve scalability \cite{DLLMMS2017}.


Current approaches for the development of server-side applications use mainstream programming languages. These languages, frequently based on the object-oriented para\-digm, provide abstractions to manage the complexity of programs and their organization into modules. However, they are designed for the creation of single executable artifacts, called monoliths. The modules of a monolith cannot execute independently, since they interact by sharing resources (memory, databases, files,\dots).
Microservices support a different view, enabling the organization of systems as collections of small \emph{independent} components. Independent refers to the capability of executing each microservice on its own machine (if needed).
This can be achieved because services have clearly defined boundaries and interact purely by means of message passing.
Microservices inherit some features from SOAs, but they take the same ideas to a much finer granularity, from programming in the large to programming in the small. Indeed, differently from SOAs, microservices highlight the importance for services to be \emph{small}, hence easily reusable, easily understood, and even easily rebuilt from scratch if needed. This recalls the single responsibility principle of object-oriented design~\cite{M03}.

Since it is convenient to abstract from the heterogeneity of possible machines (e.g., available local libraries and other details of the OS), it is useful to package a service and all its local dependencies inside a \emph{container}. Container technologies, like Docker~\cite{docker}, enable this abstraction by isolating the execution of a service from that of other applications on the same machine.
Indeed, in the literature about microservices, the emphasis is on deployment: since microservices live inside containers they can be easily deployed at different locations. A major reason for this focus is that microservices are thought since their inception as a style to program in the cloud, where deployment and relocation play key roles. Even if microservices have now evolved well beyond cloud computing, the emphasis on deployment and containerization remains~\cite{F14}.

In this chapter, while supporting the current trend of microservices, we advocate for moving the emphasis from deployment to development, and in particular to the programming language used for development.
We think that the chosen language should support the main mechanism used to build microservice architectures, namely service composition via message passing communications. Furthermore, in order to master the related complexity, we support the use of well-specified interfaces to govern communication.
Mainstream languages currently used for the development of microservices do not provide enough support for such communication modalities. In particular, service coordination is currently programmed in an unstructured and ad-hoc way, which hides the communication structure behind less relevant low-level details.

While the idea of current methodologies for developing microservices can be summarized as ``it does not matter how you develop your microservices, provided that you deploy them in containers'', the key idea behind our methodology is ``it does not matter how you deploy your microservices, provided that you build them using a microservice programming language''.
We describe our methodology in general, and support the abstract discussion by showing how our ideas are implemented in the Jolie language~\cite{MGZ14,joliewebsite}.

%
%

%

\section{Language-based Approach}
\label{sec:language}

The fine granularity of microservices moves the complexity of applications from the implementation of services to their coordination. Because of this, concepts such as communication, interfaces, and dependencies are central to the development of microservice applications. We claim that such concepts should be available as first-class entities in a language that targets microservices, in order to support the translation of the design of a microservice architecture (MSA) into code without changing domain model. This reduces the risk of introducing errors or unexpected behaviours (e.g., by wrong usage of book-keeping variables).

What are then the key ingredients that should be included in a microservice language?
Since a main feature of microservices is that they have a small size, realistic applications are composed by a high number of microservices. Since microservices are independent, the interactions among them all happen by exchanging messages. Hence, programming an MSA requires to define large and complex \emph{message exchange structures}. The key to a ``good'' microservice language is thus providing ways to modularly define and compose such structures, in order to tame complexity. We discuss such ways in the rest of this section.

{\bf Interfaces}
In order to support modular programming, it is necessary that services can be deployed as ``black boxes'' whose implementation details are hidden. However, services should also provide the means to be composed in larger systems. A standard way of obtaining this is to describe via \emph{interfaces} the functionalities that services provide to and require from the environment. Here, we consider interfaces to be sets of \emph{operations} that can be remotely invoked. Operations may be either fully asynchronous or follow the typical request-response pattern. An operation is identified by a name and specifies the data types of the parameters used to invoke it (and possibly also of the response value).

Once we accept that interfaces are first-class citizens in microservices, it makes sense to have operators to manipulate them. Since interfaces are sets of operations, it is natural to consider the usual set-theoretical operators, such as union and intersection. For example, a gateway service may offer an interface that is the union of all the interfaces of the services that it routes messages to.

{\bf Ports}
Microservices may run in heterogeneous environments that use different communication technologies (e.g., TCP/IP sockets, Bluetooth, etc.) and data protocols (e.g., HTTPS, binary protocols, etc.).
Moreover, a microservice may need to interact with many other services, each one possibly offering and/or requiring a different interface.
A communication \emph{port} concretely describes how some of the functionalities of a service are made available to the network, by specifying the three key elements above: interface, communication technology, and data protocol.
Each service may be equipped with many ports, of two possible kinds. \emph{Input ports} describe the functionalities that the service provides to the rest of the MSA. Conversely, \emph{output ports} describe the functionalities that the service requires from the rest of the MSA. Ports should be specified separately from the implementation of a service, so that one can see what a service provides and what it needs without having to check its actual implementation. This recalls the use of type signatures for functions in procedural programming, with the difference that here the environment is heterogeneous  and we thus need further information (communication medium, data protocol).
\begin{example}\label{ex:online}
Consider an online shopping service connected to both the Internet and a local intranet. This service may have 2 input ports, \lstinline+Customers+ and \lstinline+Admin+, and 1 output port, \lstinline+Auth+. Input port \lstinline+Customers+ exposes the interface that customers can use on the web, using HTTPS over TCP/IP sockets. Input port \lstinline+Admin+ exposes the administration controls of the service to the local intranet, using a proprietary binary protocol over TCP/IP sockets. Finally, output port \lstinline+Auth+ is used to access an authentication service in the local network.
\end{example}

{\bf Workflows}
Service interactions may require to perform multiple communications. For example, our previous \lstinline+Customers+ service may offer a "buy and ship" functionality, implemented as a structured protocol composed by multiple phases. First, the customer may select one or more products to buy. In the second phase, the customer sends her destination address and selects the shipment modality. Finally, the customer pays, which may require the execution of an entire sub-protocol, involving also a bank and the shipper.
Since structured protocols appear repeatedly in microservices, supporting their programming is a key issue. Unfortunately, the programming of such \emph{workflows} is not natively supported by mainstream languages, where all possible operations are always enabled. For example, consider a service implemented by an object that offers two operations \lstinline+login+ and \lstinline+pay+. Both operations are enabled at all times, but invoking \lstinline+pay+ before \lstinline+login+ raises an error. This causal dependency is programmed by using a book-keeping variable, which is set when \lstinline+login+ is called and is read by method \lstinline+pay+. Using book-keeping variables is error-prone, in particular it does not scale when the number of causality links increases~\cite{M16}.

A microservice language should therefore provide abstractions for programming workflows. For example, one can borrow ideas from BPEL~\cite{WS-BPEL}, process models~\cite{M80} or behavioral types~\cite{csurBETTY}, where the causal dependencies are expressed syntactically using operators such as sequential and parallel compositions, e.g., \lstinline+login;pay+ would express that \lstinline+pay+ becomes available only after \lstinline+login+. All the vast literature on business process modeling \cite{YanMCU07} can offer useful abstractions to this regard.

{\bf Processes}
A workflow defines the blueprint of the behavior of a service. However, at runtime, a service may interact with multiple clients and other external services. In our online shopping example, service \lstinline+Customers+ may have to support multiple users. Beyond that, the authentication service may be used both by service \lstinline+Customers+ and by other services, e.g., a \lstinline+Billing+ service.
This is in line with the principle that microservices can be reused in different contexts.
A service should thus support multiple executions of its workflow, and such executions should operate concurrently (otherwise, a new request for using the service would have to wait for previous usages to finish before being served).

A \emph{process} is a running instance of a workflow, and a service may include many processes executing concurrently. The number of processes changes at runtime, since external parties may request the creation of a new process, and processes may terminate. Each process runs independently of the others, to avoid interference, and, as a consequence, it has its own private state.

\section{The Jolie language}
\label{sec:jolie}
Jolie~\cite{MGZ14,joliewebsite} is a language that targets microservices directly. Jolie was designed following the ideas discussed in Section~\ref{sec:language}. These were initially targeted at offering a language for programming distributed systems where all components are services, which later turned out to become the microservices paradigm.

Jolie is an imperative language where standard constructs such as assignments, conditionals, and loops are combined with constructs dealing with distribution, communication, and services. Jolie takes inspiration from WS-BPEL~\cite{WS-BPEL}, an XML-based language for composing services, and from classical process calculi such as CCS~\cite{M80} (indeed, the core semantics of Jolie is formally defined as a process calculus~\cite{GLGBZ06,MC11}), but transfers these ideas into a full-fledged programming language. While we refer to~\cite{MGZ14,joliewebsite} for a detailed description of the Jolie language and its features, we discuss below the characteristics of Jolie that make it an instance of the language-based approach to microservices that we are advocating for.

The connection between Jolie and MSAs is at a very intimate level, e.g., even a basic building block of imperative languages like variables has been restructured to fit into the microservice paradigm. Indeed, microservices interact by exchanging data that is typically structured as trees (e.g., JSON or XML, supported in HTTP and other protocols) or simpler structures (e.g., database records). Thus, Jolie variables always have a tree structure~\cite{MC11}, which allows the Jolie runtime to easily marshal and unmarshal data. It is clear that such a deep integration between language and microservice technologies cannot be obtained by just putting some additional library or framework on top of an existing language (which, e.g., would rely on variables as defined in the underlying language), but requires to design a new language from the very basic foundations.

A main design decision of the Jolie language is the separation of concerns between behavior and deployment information~\cite{MGZ07,MGZ14}. Here with deployment information we mean both the addresses at which functionalities are exposed, and the communication technologies and data protocols used to interact with other services.
In particular, as discussed in the previous section, each Jolie service is equipped with a set of ports: input ports through which the service makes its functionalities available, and output ports used to invoke external functionalities.
Thanks to this separation of concerns, one can easily change how a Jolie microservice communicates with its environment without changing its behavior.  

\begin{example}
If the online shopping service of Example~\ref{ex:online} is implemented in Jolie, its \lstinline+Customers+ input port can be declared as:
\begin{lstlisting}
inputPort Customers {
  Location: "socket://www.myonlineshop.it:8000"
  Protocol: https
  Interfaces: CustomersInterface
  }
\end{lstlisting}
The port declaration specifies the location where the port is exposed, which includes both the communication technology, in this case a TCP/IP socket, and the actual URL. Also, it declares the data protocol used for communication, in this case HTTPS. Finally, the port refers to the interface used for communication, which is described separately, and which can take, e.g., the form:
\begin{lstlisting}
interface CustomersInterface { 
  RequestResponse: getList( void )( productIdList ),
                   getPrice( productId )( double ),
                   ...
  }
\end{lstlisting}
This interface provides two request-response operations, \lstinline+getList+ and \lstinline+getPrice+, and their signature. Types \lstinline+void+ and \lstinline+double+ are built-in, while \lstinline+productIdList+ and \lstinline+productId+ are user defined, hence their definition has to be provided.
\end{example}

Having ports and interfaces as first-class entities in the language allows one to clearly understand how a service can be invoked, and which services it requires. Furthermore, the same functionality can be exposed in different ways just by using different input ports, without replicating the service. For instance, in the example above, one can define a new port providing the same functionality using the SOAP protocol.
Finally, part of the port information, namely location and protocol, can be changed dynamically by the behavior, improving the flexibility. 

Jolie also supports workflows \cite{Safina2016}. Indeed, each Jolie program is a workflow: it includes receive operations from input ports and send operations to output ports, combined with constructs such as sequence, conditional and loop.
Notably, it also provides parallel composition to enable concurrency, and input-guarded choice to wait for multiple incoming messages. Input ports are available only when a corresponding receive is enabled, otherwise messages to this port are buffered.
\begin{example}
Jolie also provides novel workflow primitives that can be useful in practical scenarios. An example is the provide-until construct~\cite{M16}, which allows for the programming of repetitive behavior driven by external participants. Using standard workflow operators and provide-until, we can easily program a workflow for interacting with customers in our online shopping service from Section~\ref{sec:language}.
\begin{lstlisting}
main {
	login()( csets.sid ) { csets.sid = new };
	provide
		[ addToCart( req )( resp ) { /* ... */ } ]
		[ removeFromCart( req )( resp ) { /* ... */ } ]
	until
		[ checkout( req )( resp ) { pay; ship } ]
		[ logout() ] 
     }
\end{lstlisting}
The workflow above starts by waiting for the user to login (operation \lstinline+login+ is a request-response that returns a fresh session identifier \lstinline+sid+, used for correlating incoming messages~\cite{MC11} from the same client later on).
We then enter a provide-until construct where the customer is allowed to invoke operations \lstinline+addToCart+ and \lstinline+removeFromCart+ multiple times, until either \lstinline+checkout+ or \lstinline+logout+ is invoked. In the case for \lstinline+checkout+, we then enter another workflow that first invokes procedure \lstinline+pay+ and then procedure \lstinline+ship+ (each procedure defines its own workflow).
\end{example}

From a single workflow multiple processes are generated. Indeed, at runtime, when a message reaches a service, correlation sets~\cite{MC11} (kept in the special \lstinline+csets+ structure in the example above) are used to check whether it targets an already running process. If so, it is delivered to it. If not, and if it targets an initial operation of the workflow, a new process is spawned to manage it. Notably, multiple processes with the same workflow can be executed either concurrently, or sequentially. This last option is mainly used to program resource managers, which need to enforce mutual exclusion on the access to the resource.

\section{Conclusions and Related Work}
We made the case for a linguistic approach to microservices, and we instantiated it on the Jolie language. Actually, any general-purpose language can be used to program microservices, but some of them are more oriented towards scalable applications and concurrency (both important aspects of microservices). Good examples of the latter are Erlang~\cite{erlang} and Go~\cite{go}. Between the two, Erlang is the nearest to our approach: it has one of the most mature implementations of processes, and some support for workflows based on the actor model (another relevant implementation of actors is the Akka framework~\cite{akka} for Scala and Java). However, Erlang and Go do not separate behavior from deployment, and more concretely do not come with explicitly defined ports describing the dependencies and requirements of services. WS-BPEL~\cite{WS-BPEL} provides many of the features we described, including ports, interfaces, workflow and processes. However, it is just a composition language, and cannot be used to program single services. Also, WS-BPEL implementations are frequently too heavy for microservices. 

Our hope is that other languages following the language-based approach would emerge in the near future. This would also allow one to better understand which features are key in the approach, and which ones are just design decisions that can be changed. For instance, it would be interesting to understand whether language support for containerization would be useful, and which form it could take. Such support is currently absent in Jolie, but it would provide a better integration with the classic approach focused on deployment. A related topic would be to understand how to improve the synergy between Jolie and microservices on one side, and the cloud and IoT on the other side.

%
%

\bibliographystyle{plain}
\bibliography{MS}

\end{document}